\def\ba{\begin{eqnarray}}
\def\ea{\end{eqnarray}}
\def\l{\label}
\def\n{\nonumber \\}
\def\b{\bibitem}

\documentclass[12pt]{article}
\usepackage{epsfig}
 \begin{document} 
\title{Wounded nucleon model and Deuteron-Gold collisions at RHIC}

\author{A.Bialas and W.Czyz\\ M.Smoluchowski Institute of Physics \\Jagellonian
University, Cracow\thanks{Address: Reymonta 4, 30-059 Krakow, Poland;
e-mail:bialas@th.if.uj.edu.pl;}}
\maketitle

\begin{abstract} 
It is shown that the wounded nucleon model describes very well the
recent PHOBOS data on particle production in D-Au collisions at 200 GeV.
Contribution to particle production from a single wounded nucleon is
determined. A two-component model is formulated and shown to account for
most of the important features of the data.
\end{abstract}
 
\section {Introduction}

The model of wounded nucleons, proposed almost 30 years ago \cite{bbc}, shows a
remarkable survival capacity: it is still being used in analysis of data
\cite{wd,wb} and the very concept of a "wounded" nucleon (called now a
"participant"\footnote{According to the definition given in \cite{bbc},
the wounded nucleon is the one which underwent at least one {\it
inelastic} collision. We stick to this name because we think that the
name "participant" is incorrect: It should rather refer to nucleons
which underwent {\it any} (either elastic or inelastic) collision. })
became one of the basic tools in description and interpretation of the
heavy ion experiments. 

In its original form, the model proposes that the particle production in
a nucleus-nucleus collision can be represented as a superposition of
independent contributions from the wounded nucleons in the projectile
and in the target. Consequently, the density of particles in a collision
of nuclei of nuclear numbers $A$ and $B$ is given by

\ba
\frac{dN_{AB}}{d y} = w_A F_A(y) + w_B F_B(y)=\n=
\frac12(w_A+w_B)[F_A(y)+F_B(y)]+
 \frac12(w_A-w_B)[F_A(y)-F_B(y)]   \l{w1}
\ea
where $w_A$ and $w_B$ are the numbers of the wounded nucleons in $A$ and
$B$, $y$ is the rapidity in the c.m. system of the collision and $F_A(y)$
is a contribution from a single wounded nucleon in $A$. Similarly,
$F_B(y)$ is the contribution from a single wounded nucleon in $B$. The model
requires  
\ba
F_B(y) = F_A(-y)     \l{ww1}
\ea
but it will be convenient to keep the more general formalism.

Recently, the pseudorapidity distribution $dN/d\eta$ of particles
produced in d-Au collisions was measured by PHOBOS and BRAHMS
collaborations at RHIC in a wide range of available phase-space and for
various centralities \cite{Ph1,br}. In the present paper we use the
wounded nucleon model to analyze the data reported by PHOBOS \cite{Ph1}.

We find that the model gives a good description of the data, with the
condition (\ref{ww1}) being well satisfied, except at rapidities close
to the maximal values. This observation allows us to determine from the
data the contribution $F(\eta)$ from a single wounded nucleon.

Two novel features emerge from this analysis. It turns out that (i)
$F(\eta)$ is not confined to the hemisphere corresponding to the wounded
nucleon in question but rather extends over all available rapidity
(except possibly close to the boundary); Moreover, (ii) $F(\eta)$ shows a
distinct two-component structure.

We have shown that these observations can be understood in a recently
proposed mechanism \cite{bjez}, describing the particle production as a
two-step process: (i) multiple colour exchanges between partons from
projectile and target, followed by (ii) particle emission from
colour sources created in the first step.

In the next section we show that the wounded nucleon model describes
correctly the data from PHOBOS \cite{Ph1}. Determination of the
contributions $F_{Au} (\eta)$ and $F_D(\eta)$ from the wounded nucleons
is presented in Section 3 where also their structure is discussed. In
Section 4 we discuss a possible explanation of these findings. Our
conclusions are listed in the last section.

\section {Wounded nucleons in Deuteron-Gold collisions}

The most direct way to test the relation (\ref{w1}) is to construct
the symmetric and antisymmetric components of the particle density:
\ba
G^{\pm}(\eta)= 
\frac{dN(\eta)}{d \eta} \pm \frac{dN(-\eta)}{d \eta}  \l{1w}
\ea
In Figures 1 and 2 these two quantities are plotted versus
pseudorapidity for various centralities measured in \cite{Ph1}.
 To compare with the model, we constructed  the averages
\ba
<\Phi^{\pm}(\eta)>=  \frac{\sum_{c} G^{(c){\pm}} (\eta)}
{\sum_c[w^{(c)}_{Au}\pm w^{(c)}_D] /2}     \l{2w}
\ea
The model predicts [c.f. (\ref{w1})]
\ba
G^{\pm}(\eta)= \frac{w_{Au}\pm w_D}{2} <\Phi^{\pm}(\eta)>   \l{3w}
\ea
%rys.1
\begin{figure}[htb]
\centerline{%
\includegraphics[width=8cm]{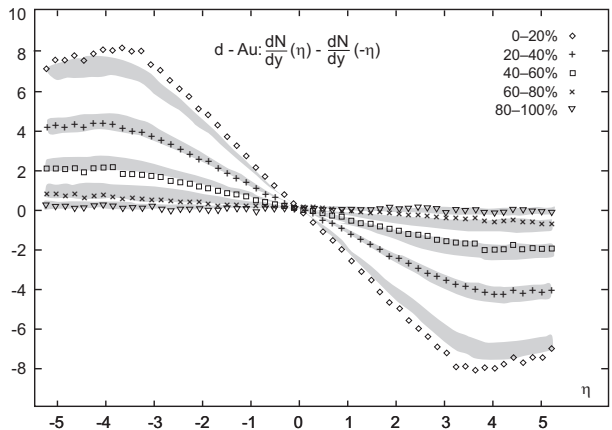}}
\caption{Antisymmetric part of the deuteron D-Au inclusive cross-section
compared with predictions of the wounded nucleon model.}
\end{figure}
%
%rys.2
\begin{figure}[htb]
\centerline{%
\includegraphics[width=8cm]{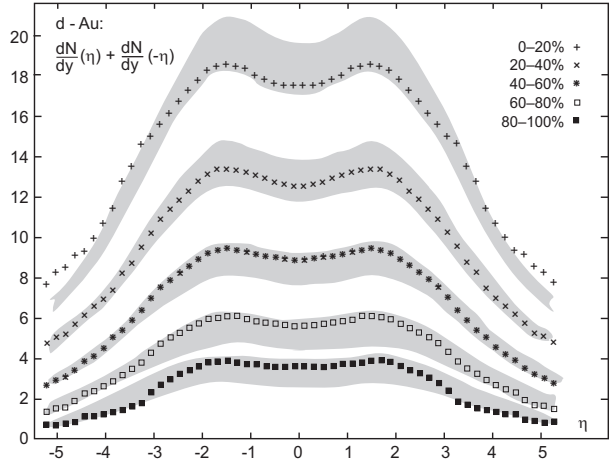}}
\caption{Symmetric part of the deuteron D-Au inclusive cross-section
compared with predictions of the wounded nucleon model.}
\end{figure}
The R.H.S of (\ref{3w}) is shown in Figs 1 and 2 as shaded areas 
(expressing the inacuracies
 in determination of $w_{Au}$ and $w_D$). One sees that the agreement is 
 rather satisfactory (except in the regions close to the maximal allowed
rapidity).

We thus conclude that the model describes correctly the available data.

Using data on nucleon-nucleon collisions another, more demanding, test of
the model is possible.

Indeed, one sees immediately from (\ref{w1}) 
that for the nucleon-nucleon collision we have
\ba
\frac{dN_{NN}}{d y} = F_N(y) +  F_N(-y)   \l{w2}
\ea
and thus for the ratio
\ba
R_{AB}(y) \equiv \frac{{dN_{AB}}/{d y}}{{dN_{NN}}/{d y}}  \l{w3}
\ea
one obtains
\ba
R_{AB}(y)= \frac12(w_A+w_B)+
 \frac12(w_A-w_B)\frac{F_A(y)-F_B(y)}{F_A(y)+F_B(y)}    \l{w4}
\ea
The first immediate consequence is 
\ba
R_{AB}(y=0)= \frac12(w_A+w_B)  \l{w5}
\ea 
implying that the value of the ratio $R_{AB}$ at mid-rapidity is
fully determined by the number of wounded nucleons and entirely 
independent of the shape of the function $F(y)$.

Fig. 3 shows $R_{D-Au}(0)$ plotted versus $(w_{Au}+w_D)/2$, as measured
by the Phobos collaboration \cite {Ph1}\footnote{The numerator of
$R_{D-Au}(0)$ was taken from the numerical data given in \cite{Ph1}. The
denominator was read off from the Figure 1b of the same paper.}. One
sees that the data are indeed in excellent agreement with (\ref{w5}). 

%rys.3
\begin{figure}[htb]
\centerline{%
\includegraphics[width=7cm]{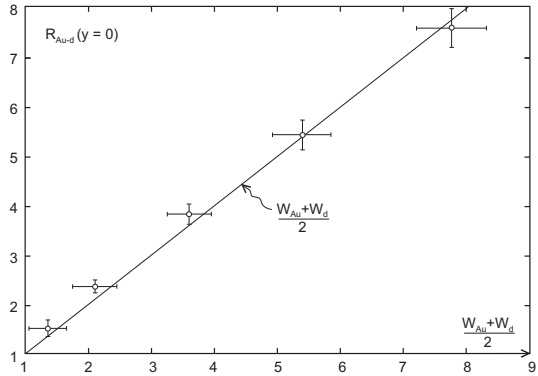}}
\caption{Particle production in the central region compared with the predictions of the wounded nucleon model.}
\end{figure}

The next step is to verify if the model gives an adequate description of
data at $y\neq 0$. To this end we propose to study the quantity
\ba
D_{D-Au}(\eta)\equiv \frac{dN_{D-Au}}{d \eta}
-R_{D-Au}(\eta=0)\frac{dN_{NN}}{d \eta}     \l{ww}
\ea
which, according to (\ref{w1}) and (\ref{w5}) should obey 
\ba
D_{Au-D}(y) = \frac12 (w_{Au}-w_D)\Phi(y)  \l{w8}
\ea
where
\ba
\Phi(y) \equiv F_{Au}(y)-F_D(y) . \l{w9}
\ea

In Fig. 4 $D_{D-Au}(\eta)$ is plotted for
various centralities, as measured in the PHOBOS experiment. 

%rys.4
\begin{figure}[htb]
\centerline{%
\includegraphics[width=8cm]{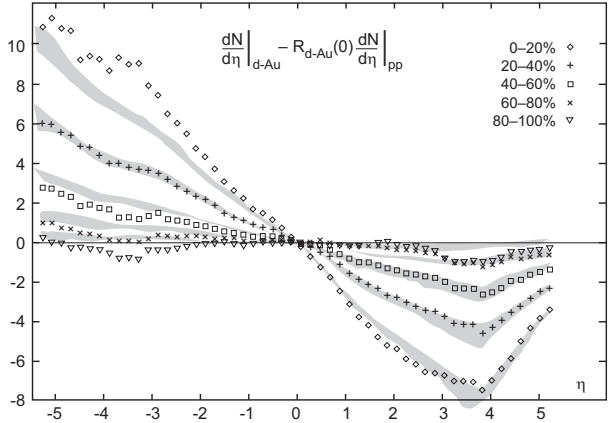}}
\caption{Comparison of the relation (10) with the predictions of the wounded nucleon model.}
\end{figure}

To verify (\ref{w8}) we again construct the ``average over centralities'': 
\ba
<{\Phi}(\eta)> = \frac{\sum_{c} D_{D-Au}^{(c)}(\eta)}
{\sum_c[w^{(c)}_{Au}-w^{(c)}_D] /2} \l{w10} 
\ea
where $c$ denotes the centrality, as measured by PHOBOS. The product
$\frac12 (w_{Au}-w_D)<{\Phi}(\eta)>$ is shown in Fig. 4 as shaded areas.
One sees good agreement with the measured values of $D_{Au-D}(\eta)$ in
the deuteron hemisphere\footnote{For maximal centrality, the approximate
linear dependence on $\eta$ was observed for $R_{D-Au}(y)$ already in
\cite{wb}. We thanks W.Busza for calling our attention to this
observations which triggered our interest in the subject.}. There are
deviations in the Au hemisphere for the most central collisions. They
may be either genuine -hitherto unexplained- deviations from the model,
or simply represent an additional contribution to particle production
from the secondary interactions inside the nucleus.  More work is needed
to clarify this feature.

\section {Particle emission  from a single wounded nucleon}

From 
(\ref{1w}) and (\ref{w1}) we deduce that the contribution from a single
wounded nucleon can be expressed as
\ba
F(\pm \eta)=\frac12[ <\Phi^+(\eta)> \pm <\Phi^-(\eta)>]        \l{r2}
\ea
 The functions $F(\pm
\eta)$ and $<\Phi^{\pm}(\eta)>)$ are shown in Fig. 5. 

%rys.5
\begin{figure}[htb]
\centerline{%
\includegraphics[width=8cm]{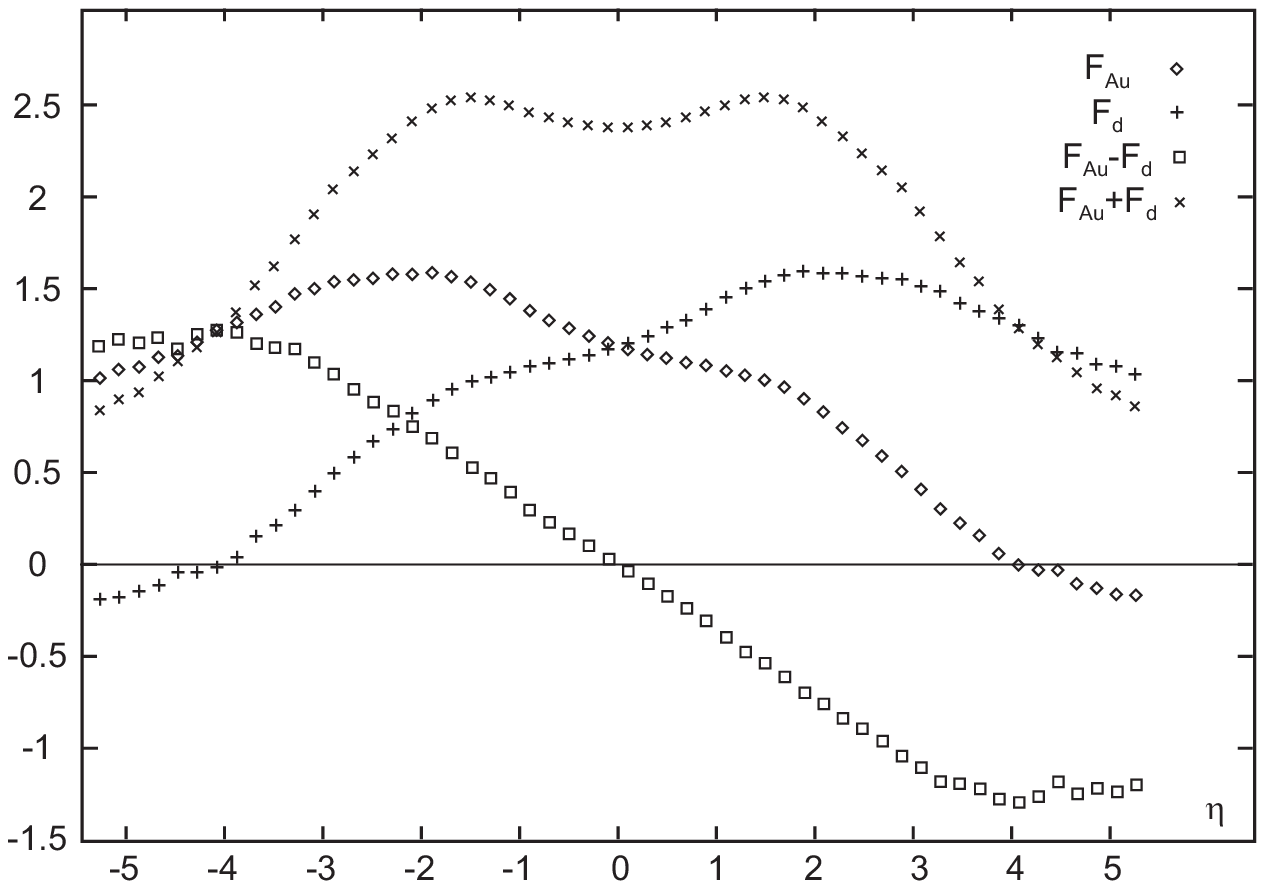}}
\caption{Particle production from a single wounded nucleon. Symmetrized particle densities.}
\end{figure}

Similarly,
using (\ref{w2}) and (\ref{w9}) and taking into account that the
symmetry relation (\ref {ww1}) is well satisfied by the data, one can
express $ \equiv F_D(\eta) \approx F_{Au}(-\eta)$, in terms of $\sigma_{pp}(\eta)
=F_D(\eta)+ F_{Au}(\eta)$ and $\bar{\Phi}(\eta)= F_{Au}(\eta)-F_D(\eta)$:

\ba
F_D=\frac12[\sigma_{pp}+\bar{\Phi}(\eta)]\;;\;\;\;
F_{Au}=\frac12[\sigma_{pp}-\bar{\Phi}(\eta)]  \l{r1}
\ea
In Fig. 6 $F_D(\eta)$ and $ F_{Au}(\eta)$ are shown in  together
with $\sigma_{pp}(\eta)$ and $<{\Phi}(\eta)>$.

%rys.6
\begin{figure}[htb]
\centerline{%
\includegraphics[width=8cm]{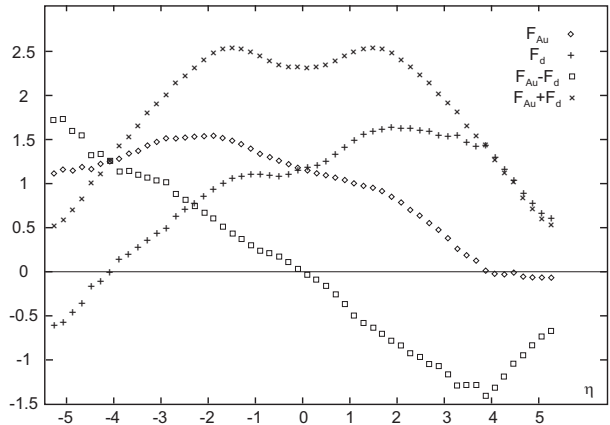}}
\caption{Particle production from a single wounded nucleon. Unsymmetrized particle densities.}
\end{figure}

One sees that, except for tiny details, the results of both
figures are very similar to each other (if one excludes the regions close
to the maximal rapidities). We thus conclude that both methods of
analysis lead to the same picture.

Three striking features are to be noted:

(a) One sees that -contrary to naive expectations-  the contribution
from a wounded nucleon extends far beyond its own hemishere, covering
practically the full rapidity interval except about 1.5 units from both
ends (where the energy conservation effects and the intranuclear cascade
are expected to give  corrections to the model in any case)
\footnote{One sees explicitely that the model does not work in the $Au$
fragmentation region, where $F_D(\eta)$ turns out negative.}. 

(b) Another -hitherto unexpected- observation is the very simple linear
dependence on $\eta$ of $<{\Phi}(\eta)>$ and of $<{\Phi}^-(\eta)>$ which can be
well approximated by a straight line with the slope of about $1/3$.

(c) There is a dramatic difference between the rapidity dependence of
the symmetric and antisymmetric part of $F(\eta)$. 

In the next section will shall discuss the consequences of these
observations for the mechanism of particle production.

\section {A possible interpretation}

The observations made in the previous section allowed us to determine the
contribution of one wounded nucleon to particle production. We have thus
obtained a qualitatively new information on mechanism of the inelastic
nucleon-nucleon collisions. A possible interpretation of this result is
presented below.

The striking difference between the measured symmetric and antisymmetric
part of $F(y)$, seen in Figs 5 and 6, suggests that $F(y)$ (which is the
sum of its symmetric and antisymmetric parts) may consists of two
components of different origin. A natural possibility is to identify one
component with particle emission from the valence part of the nucleon
and another one with emission from the gluon cloud. 

We thus write
\ba
\frac{dN}{d y}= \frac{dN^{(v)}}{d y}+\frac{dN^{(g)}}{d y}  \l{i1}
\ea
where we qualitatively expect the gluonic contribution to dominate 
the symmetric part of the spectrum, while its asymmetric part is
generated by the valence contribution.

To discuss this concept in more detail, we shall use the model proposed
recently in \cite{bjez}, thus accepting that particle production
proceeds in two-steps (i) the multi-gluon colour exchanges between
partons of the projectile and of the target and (ii) the following
emission of particle clusters from colour sources or strings.

To illustrate the consequences of this idea we shall assume that
particle emission from a colour source follows the general features of
the bremsstrahlung mechanism \cite{stod}, or -equivalently \cite{n}- the
string model \cite{lund,boa}. Consider a colour source moving to the
right. It will emit clusters\footnote{It is well known that most of the
observed particles are decay products of resonances or "clusters"
\cite{foa}.}, approximately uniformly in rapidity, until it is
neutralized by one of the partons of the target. Thus the density of the
emitted clusters, $\rho(y;y^+,y^-)$, is confined to the rapidity region
between the rapidity of the source ($y^+$) and the rapidity ($y^-$) of
this parton from the target which neutralized the source\footnote{In the
string language these are rapidities of the two ends of the string.}. 

Consequenly, the observed distribution of clusters is
\ba
\frac{dN(y)^{(v,g)}}{dy}= \int_{-Y}^y dy^- h(y^-)
\int_y^Y dy^+ H^{(v,g)}(y^+) \rho(y;y^+,y^-) \l{1i}
\ea
where 
$H^{(v,g)}(y^+)$ represent the distributions of the emitting sources 
(valence and gluons) and 
$h(y^-)$ the  distribution 
 of the partons in the target {\it normalized to 1}. Assuming that
 the parton distribution is dominated by gluons we furthermore obtain
\ba
h(y^-) = H^{(g)}(-y^-)/\int_{-Y}^Y dy H^{(g)}(y) \l{2i}
\ea

This formula immediately implies that the contribution from gluon
sources is {\it symmetric} with respect to $y$ (provided $
\rho(y;y^+,y^-)$ is symmetric, as expected). Consequently, the
contribution to the antisymmetric part of the distribution comes solely
from the valence sources.

To illustrate  other consequences of (\ref{1i}), we shall first consider
a radically simplified picture, taking $\rho(y;y^+,y^-)=\rho$ 
  for $y^-\leq y \leq y^+$, and  $H^{(g)}(y)=H^{(g)}$  
 between $-Y$ and $Y$, where $\rho$ and $H^{(g)}$ are constants. The result is
\ba
\frac{dN(y)^{(v)}}{dy} = \frac{\rho}{2Y} (Y+y)\int_y^Y dy^+ H^{(v)}(y^+)
\n
\frac{dN(y)^{(g)}}{dy}  = \frac{\rho H^{(g)}}{2Y} (Y+y)(Y-y)   \l{3i}
\ea
Since the distribution $H^{(v)}(y^+)$  of the valence part is confined
to the region close to maximal rapidity, say y$+\geq Y^*$, the integral
$\int_y^Y dy^+ H^{(v)}(y^+)$ equals 1, for $y < Y^*$.
Consequently, for $y<Y^*$ we have
\ba
\frac{dN(y)^{(v)}}{dy} = \frac{\rho}{2Y} (Y+y)     \l{4i}
\ea
i.e. the {\it linear} dependence on $y$. 

This simple exercise shows that the (observed in data) linear dependence
of the antisymmetric part of the distribution is a direct consequence of
the flat distribution of gluons and of the emitted clusters. 

The symmetric part is dominated by contribution from gluon sources which
increases linearly with the total available rapidity. The quadratic
dependence on $y$ is also -at least qualitatively- not inconsistent with
the data \cite{br}. Thus we feel that we may be indeed on the right track. 

It is clear that important refinements to this simple example are
necessary to obtain a more precise description of the data, particularly
in the region close to the maximal allowed rapidity. Some possibilities
are discussed in the Appendix.

\section {Summary and discussion}   

Using the data on pseudorapidity distributions in $D-Au$ collisions at
200 GeV c.m. energy \cite{Ph1}, we have shown that they can be
reasonably well described by the wounded nucleon model \cite{bbc}. This
allows one to determine the contribution $F(y)$ to particle production from
one wounded nucleon which is a novel information, hitherto not
available. The data show that (a) $F(y)$ extends over the full rapidity
range, far beyond the hemisphere of the wounded nucleon in question, and
(b) one observes a striking diference between the antisymmetric and
symmetric parts of $F(y)$.

The last feature suggests that $F(y)$ is built from two components,
representing particle emission from two different sources. Extending
the ideas formulated in \cite{bjez}, we proposed to identify these two
sources as (i) the valence part of the nucleon and (ii) the soft part of
the nucleon structure, dominated by gluons. This idea, accompanied with
the assumption of the approximately flat gluon rapidity spectrum,
explains immediately the striking linear behavior of the antisymmetric
part of $F(y)$, determined by the contribution from the valence source
(the gluon contribution is symmetric in rapidity and thus does not
contribute to the antisymmetric part of the spectrum).

We thus conclude that the new data on $D-Au$ collisions allowed us to
obtain a qualitatively new information on particle production and 
to identify the two distinct sources inside the nucleon.

Several comments are in order.

(i) It should be emphasized that the model of wounded nucleons  implies that
the intensity of particle emission from a wounded nucleon does not
depend on number of its interaction with the target. In our
interpretation this means that the number of colour sources per unit of
rapidity (in one nucleon) is independent of the number its interactions,
i.e., independent of the number of colour exchanges between the
projectile and target. The experimental verification of the model shows
that such a saturation is indeed present. 

(ii) It seems likely that this saturation of particle emission
is related to the concept of formation zone \cite{form}, i.e.,
strong reduction of soft emission from sources too close in rapidity. It
would be interesting to investigate this question in more detail.

(iii) Our interpretation of the data has very much in common with the dual
parton model (DPM) \cite{cap}. In particular, our "valence" contribution
corresponds to that of diquark-quark string in DPM, while our gluonic
strings are analogous to the sea-quark strings of DPM. Ignoring the
technical details (inessential at this stage of discussion), the main
difference is in the way we count the number of emitters. Although the
number of valence sources is the same in the two models, counting of the
"short" strings seems substantially different. In the dual parton model the
number of the  "short" strings equals the number of interactions between the
projectile and target. As already explained  above in  (i), in
our approach this is a property of the wounded nucleon, independent of
the number of its interaction with the target.

(iv) The simple distributions of partons and of emitted clusters used in
our discussion were taken only for illustration. If more precise
description of data is attempted, they must be accordingly modified,
particularly in the region close to maximal rapidity. One example of
possible modification is shown in the Appendix. It would be certainly
interesting to perform such an analysis when the final version of data
is available.

(v) It was shown in \cite{Ph2} that the wounded nucleon model does not
describe correctly the data for the $Au-Au$ collisions at RHIC energies.
In particular, the particle density in the central rapidity region
increases much faster than the number of wounded nucleons. It will be
very interesting to compare these deviations with the ones observed in
the present paper (c.f. Fig. 4).

\section {Appendix}

One possibility of the more adequate description of the
data, still retaining the salient features of the model, is described below. 
Let
\ba
\rho(y;y^+,y^-)= \left[1-e^{-(y^+-y)/\lambda}\right]
\left[1-e^{-(y-y^-)/\lambda}\right]\;;\n
h(y^-)=\frac{1-e^{-(y^-+Y)/\mu}}
{2Y-\mu\left[1-e^{-2 Y/\mu}\right]}\;;\n
H^{(v)}(y^+)= \frac 1{\nu}e^{-(Y-y^+)/\nu}   \l{5i}
\ea
One sees that these formulae modify the regions close to the maximal
rapidities. In the limit of vanishing  parameters $\mu$, $\nu$ and $\lambda$
 one recovers the simple situation described in Section 4. When
(\ref{5i}) is substituted into (\ref{1i}) one obtains
\ba
\frac{dN(y)^{(g)}}{dy}= \frac{W(Y+y)W(Y-y)}
{2Y-\mu\left[1-e^{-2 Y/\mu}\right]}
   \l{6i}
\ea
with
\ba 
W(z)= z -\mu\left[1+\frac{\mu}{\lambda-\mu}e^{-
z/\lambda}\right]-\lambda\left[1+\frac{\lambda}{\mu-\lambda}e^{-
z/\mu}\right] \l{7i}
\ea
and 
\ba
\frac{dN(y)^{(v)}}{dy}= \frac{W(Y+y)V(Y-y)}
{2Y-\mu\left[1-e^{-2 Y/\mu}\right]}
   \l{6ii}
\ea
where
\ba 
V(z)=1+\frac{\nu}{\lambda-\nu}e^{-(Y-y)/\nu}
+\frac{\lambda}{\nu-\lambda}e^{-(Y-y)/\lambda}  \l{7ii}
\ea
For $\lambda=\mu=\nu$ one obtains
\ba
W(z)=z\left(1+e^{- z/\mu}\right)-2\mu\left(1-e^{- z/\mu}\right)
\l{8i}
\ea
\ba
V(z)=1-\left[1+z/\nu\right]e^{-z/\nu}  \l{9i}
\ea

These formulae show explicitly that the distributions are modified in a
finite region close to the phase space boundary, where $(|Y|-|y|)$
is not too large. As one moves out from the boudary, i.e. when
$(|Y|-|y|)$ are large, the corrections vanish exponentially, and we
recover the results given by (\ref{3i}) and (\ref{4i}). The size of the
region where the corrections are important is controlled by the
parameters $\mu$, $\lambda$, and $\nu$.

We have checked that these formulae are flexible enough to account for
the results of Figs. 5 and 6.

\vspace{0.3cm} {\bf Acknowledgements} We thank Wit Busza for
illuminating discussions which focused our interest on the problem
discussed in the present paper. We also thank Krzysztof Fialkowski for
instructive discussions about different models of particle production.
Last but not least we like to thank Roman Holynski for help in dealing
with the PHOBOS data. This investigation was supported in part by the
Polish State Commitee for Scientific research (KBN) Grant No 2 P03 B
09322.

\end{document}